\documentclass[a4paper,12pt]{article}
\usepackage{empheq}
\usepackage[latin1]{inputenc}
\usepackage{float}
\usepackage{amsmath,amssymb}
\usepackage{mathrsfs}
\usepackage{theorem}
\usepackage{graphicx} 
\usepackage{color}
\usepackage{wasysym}
\usepackage{hyperref} 

\definecolor{blue}{rgb}{0,0,1}
\definecolor{green}{rgb}{0,1,0}
\definecolor{red}{rgb}{1,0,0}
\definecolor{vio}{rgb}{1,0,1}
\definecolor{ama}{rgb}{1,1,0}
\newcommand{\bc}{\begin{center}}
\newcommand{\ec}{\end{center}}
\newcommand{\be}{\nopagebreak[3]\begin{equation}}
\newcommand{\ee}{\end{equation}}
\newcommand{\ba}{\nopagebreak[3]\begin{eqnarray}}
\newcommand{\ea}{\end{eqnarray}}

{\theorembodyfont{\sffamily} }
{\theorembodyfont{\sffamily} }
{\theorembodyfont{\sffamily} }

\begin{document}

\title{\bf 
Higher dimensional conformal metrics from PDEs and Null Surface Formulation of GR 
}
\author{
Emanuel Gallo
\\
{\rm \small FaMAF, Universidad Nacional de C\'{o}rdoba,}\\
{\rm \small Instituto de F\'\i{}sica Enrique Gaviola (IFEG), CONICET,}\\
{\rm \small Ciudad Universitaria,(5000) C\'{o}rdoba, Argentina.}
}

\maketitle

\begin{abstract}

We analyze the relationship between $n$-dimensional conformal metrics and a certain class of partial differential equations (PDEs) that are in duality with the eikonal equation.
In particular, we extend the Null Surface Formulation of General Relativity to higher dimensions and give explicit expressions for the components of the metric and the generalized W\"{u}nschmann-like metricity conditions. 
We also compute the equation that the conformal factor must satisfy in order the metric be a solution of the Einstein equations.

\end{abstract}

\vspace{5mm}

\noindent

\section{Introduction}
About thirty years ago, Kozameh and Newman presented an unorthodox point of view of
General Relativity (GR) called null surface formulation (or NSF)~\cite{KN} where the dynamics 
shifted from a metric tensor on a 4-dim manifold to null surfaces
and partial differential equations (PDEs) in two dimensions. In this version of GR the conformal spacetime,
i.e., a $4$-dim manifold equipped with a conformal
structure, arises as the solution space of a particular pair of PDEs in a
$2$-dim space representing the sphere of null
directions. For the existence of this conformal structure, the pair of PDEs must satisfy the so called ``metricity conditions'', or W\"{u}nschmann conditions whose geometric meaning was unknown until recently~\cite{GKNP,Gallo2}. 

In order to carry out this approach, these authors started with a four-dimensional Lorentzian manifold, already containing a metric $g_{ab}$ and a complete integral to the
eikonal equation

\begin{equation}
g^{ab}(x^{a})\nabla _{a}Z  \nabla _{b}Z=0.  \label{EIK*}
\end{equation}
The complete integral, expressed as,
\begin{equation}
u=Z(x^{a},s ,s ^{*}),  \label{SOL*}
\end{equation}
contains the space-time coordinates, $x^{a},$ and 
two parameters or constants of integration $(s ,s ^{*}).$
By defining the four functions,
\begin{equation}
\theta ^{A}\equiv (u,\omega ,\omega ^{*},R)\equiv (Z,\partial _{s
}Z,
\partial_{s^*}Z,\partial _{s^* s }Z),  \label{THETA*}
\end{equation}
from eq.\,(\ref{SOL*}) and its derivatives, and by eliminating
$x^{a},$ via the algebraic inversion
\begin{equation}
x^{a}=X^{a}(s ,s ^{*},\theta ^{A}),  \label{INVERSION*}
\end{equation}
they found that $u=Z(x^{a},s ,s ^{*})$ satisfies in addition to
eq.\,(\ref{EIK*}) the pair of second-order partial differential
equations in $ (s ,s ^{*}),$ of the form
\begin{eqnarray}
\partial _{s s }Z & = &S (Z, \partial
_{s }Z, \partial _{s ^{*}}Z, \partial _{s s
^{*}}Z, s , s ^{*}), \label{PAIR*s}\\
\partial _{s ^{*}s ^{*}}Z & = &S^{*}(Z,
\partial _{s }Z, \partial _{s ^{*}}Z, \partial _{s s ^{*}}Z,
s , s ^{*}).\label{PAIR*}
\end{eqnarray}

The $x^{a},$ in the solution of eq.\,(\ref{PAIR*s})-(\ref{PAIR*}), appear now as
constants of integration. Therefore, the roles of $x^{a}$ and $(s,$ $s ^{*})$
are exchanged, and it is said that this system of PDEs and the eikonal equation are in duality. 
Furthermore, in this system of PDEs the metric is not present anymore. 
In fact, as was shown by Frittelli, Kozameh and Newman, a conformal metric can be reconstructed from a pairs of PDEs of the form (\ref{PAIR*s})-(\ref{PAIR*}) if ($S, S^{*})$
satisfy an integrability condition, a weak inequality and a
certain set of differential conditions $\frak{m}[S,S^*]=0,\frak{m}^*[S,S^*]=0$, (the metricity or
generalized W\"{u}nschmann conditions), i.e. the conformal structure is codified in $Z$ or alternatively in the functions $S$ and $S^*$.

The solutions to the pair determine a conformal
four-dimensional Lorentzian metric and, in fact, all conformal
Lorentzian metrics can be obtained from equivalence classes of
equations of the form eq.(\ref{PAIR*s})-(\ref{PAIR*}). When certain specific
conditions\,\cite{KNN}, in addition to the W\"{u}nschmann
condition, are imposed on the ($S ,S^{*}),$ the
metrics, determined by the solutions, are in the conformal
Einstein vacuum class.\newline
The purpose of this work is to show that a similar program to the null surface formulation of general relativity can be developed in dimensions higher than three and four~\cite{KN}-\cite{Forni}. Analogous results were presented in~\cite{Gallo3} for the Hamilton-Jacobi formulation of General Relativity from Montiel-Pi$\tilde{\rm n}$a, Newman and Silva-Ortigoza~\cite{ETG}.
In particular, we show that from a special system of PDEs,(conformal) Lorentzian metrics can always be constructed , which are in duality with the eikonal equation.

In order to achieve this objective, we will use the same techniques as were developed by Frittelli, Kozameh and Newman ~\cite{KN}-\cite{ FCN3} in order to obtain NSF in its $4$-dim version. 

We will present the essential results, and we will see how the imposition of the Einstein equations to the system determine a unique $n$-dim metric.

\section{ The duality between the eikonal equation and a system of 2nd order PDEs}

Let ${\cal M}$ be an $n$-dimensional manifold  with local
coordinates $x^{a}=(x^{0},...,x^{(n-1)})$ and let us assume that we are given
a ($n-2$)-parameter set of functions $u=Z(x^{a}, s, s^*,
\gamma^m)$, with $m = 1,...,(n-4)$. 

The parameters $s$, $s^*$ and
$\gamma^m$ can take values on an open neighborhood of a manifold
${\cal N}$ of dimension ($n-2$). It will be also assumed that for fixed
values of the parameters $s$, $s^*$ and $\gamma^m$ the level
surfaces

\begin{equation}
u = constant = Z(x^{a}, s, s^*, \gamma^m),  \label{sol2*}
\end{equation}
locally foliate the manifold $\mathcal{M}$ and that $Z(x^{a}, s,
s^*, \gamma^m)$ satisfies the eikonal equation
\begin{equation}
g^{ab}(x^{a})\nabla _{a}Z(x^{a}, s, s^*, \gamma^m)\nabla
_{b}Z(x^{a}, s, s^*, \gamma^m)=0,  \label{Eik2}
\end{equation}
for some Lorentzian metric $g_{ab}(x^{a})$.

Therefore, for each fixed value of $\{s,s^*,\gamma^m\}$, the level surfaces $Z(x^a,s,s^*,\gamma^m)=\text{constant}$, are null surfaces of $({\cal M},g_{ab})$.

We want to find now, a system of PDEs in duality with the eikonal equation, i.e. a system that admits the same solutions, but where the role of integration constants and parameters is exchanged.

From the assumed existence of $Z(x^{a}, s, s^*, \gamma^m)$, we
define $n$ parameterized scalars $\theta ^{A}$, with ($A = 0, +,
-, m,R$), in the following way:
\begin{eqnarray}
\theta ^{0} &=& u =Z, \label{tetaA1} \\
\theta ^{+} & = &  w^+ =\partial_sZ,\\
\theta ^{-} & = & w^- =\partial_{s^*}Z, \\
\theta ^{m} & = & w^m=\partial_mZ,\\
\theta ^{R} & = & R =\partial_{ss^*}Z,\label{tetaA5}
\end{eqnarray}\\
where the derivatives with respect to the parameters $s$, $s^*$, and $\gamma^m$ are denoted by  $\partial_{s}$,
$\partial_{s^*}$ and $\partial_{\gamma^m} = \partial_m$. In a similar way, differentiation with respect to the local coordinates $x^a$, will be denoted as $\nabla_a$ or ``comma a,'', and for an arbitrary function $F(\theta^A,s,s^*,\gamma^m)$, $F_{\theta^A}$ will be the partial derivative of $F$ with respect to $\theta^A$.

We will assume that $Z(x^{a}, s, s^*, \gamma^m)$ is such that eqs.\thinspace (\ref{tetaA1})-(\ref{tetaA5})
can be solved for the $x^{a}$'s for all values of $\{s,s^*,\gamma^m\}$ in an open neighborhood $\cal {O} \in \cal {N}$; that is, we require
\begin{equation}
\text{det}\;\theta^A{}_{,b}\neq 0;\label{eq:inverse}
\end{equation}
and therefore
\begin{equation}
x^{a}=X^{a}(u,w^+, w^-, w^m,R, s ,s^*,\gamma^m ).\label{coordinad}
\end{equation}

It can be shown that in the case of flat Lorentzian spacetimes there exist families of
null surfaces where eq.(\ref{eq:inverse}) is satisfied, (see appendix $A$ for a discussion). Therefore, it follows the existence on arbitrary spacetimes of (local) families of null surfaces where eq.(\ref{eq:inverse}) is also satisfied.

Assuming this, let us note that for each fixed value of $s$, $s^*$ and
$\gamma^m$ eqs.\,(\ref{tetaA1})-(\ref{tetaA5}) can be thought as a coordinate
transformation between the $x^a$'s and the $\theta^A$'s.

Defining the following $n(n-3)/2$ scalars
\begin{eqnarray}
\tilde S(x^a, s,s^*,\gamma^m) &=& \partial_{s s}Z(x^{a}, s, s^*, \gamma^m), \label{tildef1} \\
\tilde {S}^*(x^a, s,s^*,\gamma^m) &=& \partial_{s^* s^*}Z(x^{a}, s, s^*, \gamma^m), \\
\tilde{\Phi}_{m}(x^a, s,s^*,\gamma^m) &=& \partial_{s m}Z(x^{a}, s, s^*, \gamma^m),\\
\tilde{\Phi}^*_{m}(x^a, s,s^*,\gamma^m) &=& \partial_{s^* m}Z(x^{a}, s, s^*, \gamma^m), \\
\tilde{\Upsilon}_{lm}(x^a, s,s^*,\gamma^m) &=& \partial_{l m}
Z(x^{a},s, s^*, \gamma^m),\label{tildef2}
\end{eqnarray}\\
and taking into account eq.(\ref{coordinad}), we obtain a system of PDE's dual to the eikonal equation given
by

\begin{eqnarray}
\partial_{s s}Z & = & S_{}(u,w^+,w^-, w^m,R,s,s^*,\gamma^m),\label{PDEIIa} \\
\partial_{s^* s^*}Z & = & S^*_{}(u,w^+,w^-, w^m,R,s,s^*,\gamma^m), \\
\partial_{s m}Z & = & \Phi_{m}(u,w^+,w^-, w^m,R,s,s^*,\gamma^m),\\
\partial_{s^* m}Z & = & \Phi^*_{m}(u,w^+,w^-, w^m,R,s,s^*,\gamma^m), \\
\partial_{l m}Z & = &\Upsilon_{lm}(u,w^+,w^-, w^m,R,s,s^*,\gamma^m);\label{PDEIIb}
\end{eqnarray}
where \begin{equation}
S_{}(u,w^+,w^-, w^m,R,s,s^*,\gamma^m)=\tilde S_{}(x^a(u,w^+,w^-, w^m,R,s,s^*,\gamma^m),s,s^*,\gamma^m),
\end{equation} 
and so on.

It means that the ($n-2$)-parametric family of level surfaces
eq.\thinspace (\ref{sol2*}), can be obtained as solutions of the $n(n-3)/2$ system of second order PDEs (\ref{PDEIIa})-(\ref{PDEIIb}). In this case
 ($S_{},S^*_{},\Phi_{m},
\Phi^*_{m},\Upsilon_{lm} $) satisfy the following integrability conditions:
\begin{eqnarray}
D_k S&= &D_s\Phi_k,\\
D_k S^*&=& D_{s^*}\Phi^*_k,\\
D_k  \Phi_m&= &D_m\Phi_k=D_{s}\Upsilon_{mk},\\ 
D_k  \Phi^*_m&=&D_m\Phi^*_k=D_{s^*}\Upsilon_{mk},\\
D_i\Upsilon_{mk}&=&D_k\Upsilon_{mi},\\
D_sT^*&=&D_{s^*}T,\\
D_mQ_k&=&D_kQ_m,\\
D_mT&=&D_{s}Q_m,\\
D_mT^*&=&D_{s^*}Q_m;
\end{eqnarray}
where:\\\\

\textbf{Definition 1:} \textit{The total $s$, $s^*$ and $\gamma^m$
derivatives of a function $F=F(\theta^A, s, s^*,
\gamma^n)$ are defined by}
\begin{eqnarray}
D_{s}F &\equiv &\partial_{s}F + F_{u} w^+ + F_{w^+} S_{} +
F_{w^-} R + F_{R} T + F_{w^m} \Phi_{m}, \\
D_{s^*}F &\equiv &\partial_{s^*}F + F_{u} w^- + F_{w^-} S^*_{} +
F_{w^+} R + F_{R} T^* + F_{w^m} \Phi^*_{m}, \\
D_{m}F &\equiv &\partial_{m}F + F_{u} w^m + F_{w^+} \Phi_{m} + F_{w^-}
\Phi^*_{m} + F_RQ_m +F_{w^k} \Upsilon_{km}, \label{totalND}
\end{eqnarray}
\textit{respectively, with}
\begin{eqnarray}
T & = & \frac{1}{1 - S_{R} S^*_{R}}\Big[S_{s^*} + S_{u} w^- +
S_{w^-} S^*_{} + S_{w^+} R +
S_{w^m} \Phi^*_{m} \nonumber \\ 
& &+ S_{R}\Big(S^*_{s} +
S^*_{u} w^+ + S^*_{w^+} S_{} 
+ S^*_{w^-}R
+ S^*_{w^m} \Phi_{m}\Big)\Big],\label{T} \\
 T^*& = & \frac{1}{1 - S_{R} S^*_{R}}\Big[S^*_{s} + S^*_{u} w^+ +
S^*_{w^+} S_{} + S^*_{w^-} R+
S^*_{w^m} \Phi_{m}\nonumber \\ 
& & + S^*_{R}\Big(S_{s^*} +
S_{u} w^- + S_{w^-} S^*_{} + S_{w^+}R
+ S_{w^m} \Phi^*_{m}\Big)\Big],\label{Tstar} \\
Q_m & = & \Phi_{m,s^*} + \Phi_{m,u} w^- + \Phi_{m,w^-}
S^*_{} +\Phi_{m,w^+} R + \Phi_{m,R} T^*
+ \Phi_{m,w^k} \Phi^*_{k}. \label{Qm}
\end{eqnarray}\\

Note that
\begin{eqnarray}
 T&\equiv & D_sR=D_{s^*}S,\\ 
 T^*&\equiv &D_{s^*}R=D_{s}S^*,\\
 Q_m&\equiv &D_mR=D_{s^*}\Phi_m=D_{s}\Phi^*_m.
 \end{eqnarray}
It is easy to show that if the functions ($S_{},S^*_{},\Phi_{m},\Phi^*_{m},\Upsilon_{lm} $) satisfy the  integrability conditions, the solution space of eqs.(\ref{PDEIIa})-(\ref{PDEIIb}) is $n$-dimensional; as we show next.

The system of PDE's (\ref{PDEIIa})-(\ref{PDEIIb}) is equivalent to the Pfaffian system generated by the $n$
one-forms, $\beta^A = (\beta^0, \beta^+, \beta^-, \beta^m,\beta^R)$
\begin{eqnarray}
\beta^{0} & =& du - w^+ ds - w^- ds^* - w^m d \gamma^m,  \label{Frob} \\
\beta^{+} & =&dw^+ - Sds - R ds^* - \Phi_m\, d\gamma^m,  \\
\beta^{-} & =&dw^- - R ds - S^* ds^* - \Phi^*_{m}\,d\gamma^m, \\
\beta^{m} & =&dw^m - \Phi_m ds - \Phi^*_m ds^* -\Upsilon_{mk}\,d \gamma^k,\\
\beta^R & =&dR - T ds - T^*ds^* -Q_{k}\, d \gamma^k.
\end{eqnarray}

A direct computation shows that,
\begin{eqnarray}
d\beta^{0} & = & ds\wedge\beta^+ +ds^*\wedge\beta^- +d\gamma^k\wedge\beta^k, \\
\nonumber \\
d\beta^{+} & = &[S_{u} ds + \Phi_{k,u} d \gamma^k]
\wedge \beta^{0} + [S_{w^+} ds + \Phi_{k,w^+} d
\gamma^k] \wedge \beta^{+} \nonumber \\ & & 
+ [S_{w^-} ds
+ \Phi_{k,w^-} d \gamma^k] \wedge \beta^{-}  +
[S_{w^k}
ds + \Phi_{m,w^k} d \gamma^m] \wedge \beta^{k} \nonumber \\
& & 
+[S_{R}
ds + \Phi_{k,R} d \gamma^k+ds^*] \wedge \beta^R \nonumber \\
& & 
+ [D_k
S_{} - D_s \Phi_{k}]ds \wedge d\gamma^k  
- D_k\Phi_{m} d\gamma^k \wedge d\gamma^m,\\
\nonumber \\
d\beta^{-} & = &[S^*_{u} ds^* + \Phi^*_{k,u} d \gamma^k]
\wedge \beta^{0} + [S^*_{w^-} ds^* + \Phi^*_{k,w^-} d
\gamma^k] \wedge \beta^{-} \nonumber \\ & & + [S^*_{w^+}
ds^* + \Phi^*_{k,w^+} d \gamma^k] \wedge \beta^{+}  +
[S^*_{w^k}
ds^* + \Phi^*_{m,w^k} d \gamma^m] \wedge \beta^{k} \nonumber \\
& & +
[ds+S^*_{R}
ds^* +\Phi^*_{k,R} d \gamma^k] \wedge \beta^R 
+ [D_kS^*_{} - D_{s^*} \Phi^*_{k}]ds^* \wedge d\gamma^k   
- D_m\Phi^*_{k} d\gamma^k \wedge d\gamma^m,\\
\nonumber \\
d\beta^{m} & = & [\Phi_{m,u} ds + \Phi^*_{m,u} ds^* +
\Upsilon_{mk,u} d \gamma^k] \wedge \beta^{0} \nonumber \\ & & +
[\Phi_{m,w^+} ds + \Phi^*_{m,w^+} ds^* + \Upsilon_{mk,w^+} d
\gamma^k] \wedge \beta^{+} 
+ [\Phi_{m,w^-} ds + \Phi^*_{m,w^-} ds^* + \Upsilon_{mk,w^-}
d \gamma^k] \wedge \beta^{-} \nonumber \\
&& + [\Phi_{m,w^j} ds + \Phi^*_{m,w^j} ds^* + \Upsilon_{mk,w^j}
d \gamma^k] \wedge \beta^{j} 
+ [\Phi_{m,R} ds + \Phi^*_{m,R} ds^* +\Upsilon_{mk,R} d \gamma^k] \wedge \beta^{R} \nonumber \\ & & 
+[D_i \Phi^*_{m}-D_{s^*} \Upsilon_{mi} ]ds^* \wedge d\gamma^i  
+ [D_i \Phi_{m}-D_{s} \Upsilon_{mi}]ds \wedge d\gamma^i -
D_i \Upsilon_{mk} d\gamma^i \wedge d\gamma^k ,\\
\nonumber \\
d\beta^R & = & [T_{u} ds + T^*_u ds^* +
Q_{k,u} d \gamma^k] \wedge \beta^{0} +
[T_{w^+} ds + T^*_{w^+} ds^* + Q_{k,w^+} d
\gamma^k] \wedge \beta^{+} \nonumber \\
& &
 + [T_{w^-} ds + T^*_{w^-} ds^* + Q_{k,w^-}
d \gamma^k] \wedge \beta^{-}  + [T_{w^j} ds + T^*_{w^j} ds^* + Q_{k,w^j}
d \gamma^j] \wedge \beta^{k} \nonumber \\
& & + [T_{R} ds + T^*_{R} ds^* + Q_{k,R}
d \gamma^k] \wedge \beta^R \nonumber \\
& & + [D_{s^*}T - D_{s}T^*]ds \wedge ds^* +
[D_i T^*-D_{s^*} Q_i]ds^* \wedge d\gamma^i  \nonumber \\
& & + [D_i T-D_{s} Q_{i}]ds \wedge d\gamma^i -
D_i Q_{k} d\gamma^i \wedge d\gamma^k,
\end{eqnarray}
where in order to perform these computations we have used the fact that for an
arbitrary function\\ $\Lambda=\Lambda(u, w^+, w^-, R, s, s^*, \gamma^m)$, 
\begin{equation}
d\Lambda=\Lambda_u\beta^0+\Lambda_{w^+}\beta^+ +\Lambda_{w^-}\beta^-+\Lambda_{w^m}\beta^m +\Lambda_R\beta^R+D_s\Lambda\,ds+D_{s^*}\Lambda\,ds^*+D_m\Lambda\,d\gamma^m.
\end{equation}
Therefore, using integrability conditions, $d \beta^A = 0$ \,(modulo $\beta^A$).
From this result and the Frobenius theorem, we conclude that the
solution space of eqs.\thinspace(\ref{PDEIIa})-(\ref{PDEIIb}) is
$n$-dimensional.

In this way, we have obtained a system of differential equations in duality with the eikonal equation, with the metric disappeared from these equations. The natural question is as follows: Could one start with this system of
PDEs, and then find the eikonal equation, with a metric $g^{ab}(x^a)$? As in $3$ and $4$ dimensions, we will show that when the functions ($S_{},S^*_{},\Phi_{m},
\Phi^*_{m},\Upsilon_{lm} $) satisfy the integrability conditions, and a
 set of differential conditions (the metricity or generalized W\"{u}nschmann conditions), the
procedure can be reversed. The solutions of the system determine a conformal n-dimensional
metric.

In fact, there exist several geometrical ways to study these equations. One could, for example, study the equivalence problem associated to these equations, asking for a class of PDEs that can be obtained from the original one from the so called point or contact  transformations (\cite{FKamN,SCT, Gallo2,Nur1,Nur2}); or by direct construction of a conformal connection with vanishing torsion tensor.
However, there exists a more straightforward method to reconstruct the metric from $Z$ (or ($S_{},S^*_{},\Phi_{m},
\Phi^*_{m},\Upsilon_{lm} $)), and it is the method that was developed by Frittelli, Kozameh and Newman in order to obtain the so called null surface formulation of general relativity. 
Due to the simplicity of this technique, we extend its use to higher dimensions. 

\section{$n$-dimensional conformal metrics}
The basic idea now is to solve eq.\thinspace (\ref{Eik2}) for the
components of the metric in terms of $Z(x^{a}, s, s^*, \gamma^m)$.
To do so, we will consider a number of parameter derivatives of
the condition (\ref{Eik2}), and then by manipulation of these
derivatives, obtain both the $n$-dimensional metric and the conditions that the
$n(n-3)/2$ partial differential equations defining the surfaces
must satisfy. 
From the $n$ scalars, $\theta ^{A}$, we have their associated
gradient basis $\theta ^{A}{}_{\,a}$ given by
\begin{eqnarray}
\theta ^{A}{}_{\,a}=\nabla _{a}\theta ^{A} & = & \{Z_{\,a},
D_{s} Z_{\,a}, D_{s^* }Z_{\,a},D_{m}Z_{\,a},
D_{s s^* }Z_{\,a}\}
\end{eqnarray}
and its dual vector basis $\theta _{B}\,^{a}$, so that
\begin{equation}
\theta _{A}\,^{a}\theta ^{B}{}_{\,a}=\delta
_{A}\,^{B},\,\,\,\,\theta _{A}\,^{a}\theta ^{A}{}_{b}=\delta
_{b}\,^{a}.  \label{vdv}
\end{equation}

As was shown in the original works on NSF, it is easier to search for the components of the $n$-dimensional
metric in the gradient basis rather than in the original
coordinate basis. Furthermore, it is preferable to use the
contravariant components rather than the covariant components of
the metric; that is, we want to determine
\begin{equation}
g^{AB}(x^{a},s, s^*, \gamma^m)=g^{ab}(x^{a})\theta
^{A}{}_{\,a}\theta _{\,b}^{B}. \label{gijII}
\end{equation}

The metric components and the W\"{u}nschmann-like conditions are
obtained by repeatedly operating with $D_{s}$, $D_{s^*}$ and
$D_{m}$ on eq.\thinspace (\ref{Eik2}), that is, by definition, on
\begin{equation}
g^{00}=g^{ab}Z_{\,a}Z_{\,b}=0.  \label{h-j3zeta}
\end{equation}

Applying $D_{s}$ to eq.\thinspace (\ref{h-j3zeta}) yields
$D_{s}{g^{00}}=2g^{ab}\partial _{s}Z,_{a}Z,_{b}=0,$ i.e.,
\begin{equation}
g^{+0}=0.  \label{g+0}
\end{equation}

In the same way we obtain from $D_{s^* }g^{00}=0$ and $D_{m}g^{00}=0$, that
\begin{eqnarray}
g^{-0} & = & g^{m0}= 0. \label{g-0} 
\end{eqnarray}

Computing the second derivative $D_{ss}(g^{00}/2)=0$ we obtain

\begin{eqnarray}
D_{ss}(g^{00}/2)&=&g^{ab}(x^a)D_{ss}Z,_{a}Z,_{b}+g^{ab}D_{s}Z,_{a}D_{s}Z,_{b}=0\nonumber\\
&=&S_{R}g^{0R}+g^{++}=0, 
\end{eqnarray}
and therefore
\begin{equation}
g^{++}= -S_{R} g^{0R}.
\end{equation}

Similarly applying the second derivatives $D_{s^*s^*},D_{ss^*},D_{ms},D_{ms^*}$ and $D_{nm}$ to $g^{00}$ yields
\begin{eqnarray}
g^{--}&=& -S^*_{R} g^{0R},\\
g^{+-}&=&-g^{0R},\\
g^{+m}&=&-\Phi_{m,R}g^{0R},\\
g^{-m}&=&-\Phi^*_{m,R}g^{0R},\\
g^{nm}&=&-\Upsilon_{nm,R}g^{0R};
\end{eqnarray}
where in all these computations it
we used the fact that for an
arbitrary function $F(\theta^A, s, s^*, \gamma^m)$ one has $F,_{a}=F_{\theta^A}\theta ^{A}{}_{\,a}$.

From the third derivatives
\begin{eqnarray}
D_{sss^*}(g^{00}/2)&=&g^{ab}(x^a)T^*{},_{a}u,_{b}+g^{ab}(x^a)S^*,_{a}w^+{},_{b}+2g^{-R}=0,\\
D_{s^*ss^*}(g^{00}/2)&=&g^{ab}(x^a)T,_{a}u,_{b}+g^{ab}(x^a)S,_{a}w^-{},_{b}+2g^{+R}=0,
\end{eqnarray} we obtain a linear system for $g^{+R}$ and $g^{-R}$ that can be solved if $S_{R}S_{R}^*\neq 4$; then we find

\begin{eqnarray}
g^{+R}&=&-\frac{g^{0R}}{4-S_{R}S^*_{R}}
[2(T_R-S_{w^+}-S_{w^-}S^*_{R}
-S_{w^m}\Phi^*_{m,R})\nonumber\\
&&-S_{R}(T^*_R-S^*_{w^-}-S^*_{w^+}S_{R}
-S^*_{w^m}\Phi_{m,R})], 
\end{eqnarray}
and
\begin{eqnarray}
g^{-R}&=&-\frac{g^{0R}}{4-S_{R}S^*_{R}}
[2(T^*_R-S^*_{w^-}-S^*_{w^+}S_{R}
-S^*_{w^m}\Phi_{m,R})\nonumber\\
&&-S^*_{R}(T_R-S_{w^+}-S_{w^-}S^*_{R}
-S_{w^m}\Phi^*_{m,R})].
\end{eqnarray}

By computing 
\begin{equation}
D_{mss^*}(g^{00}/2)=g^{ab}(x^a)Q_{m,a}u,_{b}+g^{ab}(x^a)R,_{a}w^m{},_{b}+g^{ab}(x^a)\Phi^*_{m,a}w^+{},_{b}+g^{ab}(x^a)\Phi_{m,a}w^-{},_{b}=0,
\end{equation} 
it is found  
\begin{eqnarray}
g^{mR}&=&-\Phi^*_{m,R}g^{+R}- \Phi_{m,R}g^{-R}-g^{0R}(Q_{m,R}
-\Phi_{m,w^+}-\Phi_{m,w^-}S^*_{R}\nonumber\\
&&-\Phi_{m,w^n}\Phi^*_{n,R}-\Phi^*_{m,w^+}S_{R}
-\Phi^*_{m,w^-}-\Phi^*_{m,w^n}\Phi_{n,R}).
\end{eqnarray}

Finally, from 
\begin{eqnarray}
D_{ss^*ss^*}(g^{00}/2)&=&g^{ab}(x^a)U,_{a}u,_{b}+2g^{ab}(x^a)T^*{},_{a}w^+{},_{b}+2g^{ab}(x^a)T,_{a}w^-{},_{b}\nonumber\\&&+2g^{RR}+g^{ab}(x^a)S,_{a}S^*,_{b}=0,
\end{eqnarray} 
with $U=D_{s^*}T$, we obtain the last metric component $g^{RR}$ (if $S_{R}S^*_{R}\neq -2$),
\begin{eqnarray}
g^{RR}&=&-\frac{1}{2+S_RS^*_R}\left\{S^*_{w^m}S_Rg^{mR}+\left[U_R-2T_{w^+}
-2T_{w^-}S^*_{R}-2T_{w^m}\Phi^*_{m,R}-2T^*_{w^+}S_{R}\right.\right.\nonumber\\
&&\left.\left.-2T^*_{w^-}-2T^*_{w^m}\Phi_{m,R}+S^*_{u}S_{R}+S^*_{R}S_{u}-S^*_{w^+}\left(S_{w^+}S_{R}+S_{w^-}+S_{w^m}\Phi_{m,R}\right)\right.\right.\nonumber\\
&&\left.\left.-S^*_{w^-}\left(S_{w^+}+S_{w^-}S^*_{R}+S_{w^m}\Phi^*_{m,R}\right)-S^*_{w^m}\left(S_{w^+}\Phi_{m,R}+S_{w^-}\Phi^*_{m,R}+S_{w^n}\Upsilon_{mn,R}\right)\right]g^{0R}\right.\nonumber\\
&&\left.+\left(2T^*_R+S^*_{w^+}S_R+S^*_RS_{w^+}\right)g^{+R}+\left(2T_R+S_{w^-}S^*_R+S_RS^*_{w^-}\right)g^{-R}\right\}.
\end{eqnarray}
The metric then is expressed as:
\begin{equation}\label{gAB}
g^{AB}=\Omega^2\frak{g}^{AB}=\Omega^2
\left(%
\begin{array}{cccccc}
  0 & 0 & 0 & ... & 0 & 1 \\
  0 & -S_{R} & -1 & ... & -\Phi_{m,R} & \frak{g}^{+R} \\
  0 & -1 & -S^*_{R} & ... & -\Phi^*_{m,R} & \frak{g}^{-R} \\
  \vdots & \vdots & \vdots & [-\Upsilon_{nm,R}] & \vdots & \vdots \\
  0 & -\Phi_{m,R} & -\Phi^*_{m,R} & ... & -\Upsilon_{mm,R} & \frak{g}^{mR} \\
  1 & \frak{g}^{+R} & \frak{g}^{-R} & \ldots & \frak{g}^{mR} & \frak{g}^{RR} \\
\end{array}%
\right) .
\end{equation}
Therefore, we have found that all contravariants components of the metric can be expressed in term of derivatives of the functions $(S,S^*,\Phi_m,\Phi^*_m,\Upsilon_{ml})$ and a conformal factor 
$g^{0R}=\Omega^2$.

It is worthwhile to mention that by construction the metric obtained in eq.(\ref{gAB}) has Lorentzian signature. However, if one starts from a system of PDEs that satisfy the W\"{u}nschmann conditions that we will present in the next section, and wishes to restrict the type of metrics to those with Lorentzian signature, one must impose extra conditions on ($S_{},S^*_{},\Phi_{m},\Phi^*_{m},\Upsilon_{lm} $). For example in 4-dim, such a condition reads $1-S_{R}S^*_{R}>0$, that follows from require $\text{det}(g^{ab})<0$, but unfortunately, in higher dimensions this last condition is not sufficient. In these cases, for each particular set of PDEs, one must to study the eigenvalue problem associated to the metric, and count the number of positive and negative signs\cite{cartanmetric,coll}.

\section{The generalized W\"{u}nschmann or metricity conditions}
If we compute the remaining third derivatives, we obtain
relations that automatically satisfies $Z$, but if we start with
the point of view that we want to construct a conformal metric
from the system of PDEs, then these relations are converted in
W\"{u}nschmann-like conditions that must satisfy our system to assure
the existence of the conformal metric in the solution space. These
conditions read, (following by applying $D_{sss}, D_{s^*s^*s^*},
D_{kmn}, D_{s^*s^*m}, D_{ssm}, D_{smn}, D_{s^*nm}$ to $g^{00}$),
\begin{eqnarray}
\frak{m}&=&D_s[S_{}\cdot u]+2[S_{}\cdot w^+]=0,\label{W1}\\
\frak{m}^*&=&D_{s^*}[S^*_{}\cdot u]+2[S^*_{}\cdot w^-]=0,\\
\frak{m}_{kmn}&=&D_k[\Upsilon_{mn}\cdot u]+[\Upsilon_{km}\cdot w^n]+[\Upsilon_{kn}\cdot w^m]=0,\\
\frak{m}_{m}&=&D_m[S_{}\cdot u]+2[\Phi_{m}\cdot w^+]=0,\\
\frak{m}^*_{m}&=&D_m[S^*_{}\cdot u]+2[\Phi^*_{m}\cdot w^-]=0,\\
\frak{m}_{mn}&=&D_s[\Upsilon_{mn}\cdot u]+[\Phi_{m}\cdot w^n]+[\Phi_{n}\cdot w^m]=0,\\
\frak{m}^*_{mn}&=&D_{s^*}[\Upsilon_{mn}\cdot u]+[\Phi^*_{m}\cdot
w^n]+[\Phi^*_{n}\cdot w^m]=0\label{W2},
\end{eqnarray}
where we are using the notation $F\cdot G=g^{ab}F,_aG,_b$, for arbitrary functions $F$ and $G$.\\
Explicit expressions for these metricity conditions are found in the appendix B.
\newline

\textbf{Remark 2:} \textit{Due to the fact that $\frak{m}_{mn}=\frak{m}_{nm}$ and $\frak{m}_{kmn}=\frak{m}_{(knm)}$, there are in total $\frac{1}{6}(n^2-4)(n-3)$ independent conditions in $n$ dimensions.
If we continue applying higher order derivatives to $g^{00}$ as it
happens in the 4-dim version of NSF, we do not obtain new
information. From these derivatives we obtain only identities
from the previous relations.}\\\\

On the other hand, if we apply the derivatives 
$D_s,D_{s^*},D_m$ to the component
$g^{0R}=\Omega^2$ of the metric, i.e., to the conformal factor, we obtain the equations
\begin{eqnarray}
D_s\Omega&=&\frac{1}{2}\left(\frak{g}^{+R}+T_R\right)\Omega,\label{Omega_relacion1}\\
D_{s^{*}}\Omega&=&\frac{1}{2}\left(\frak{g}^{-R}+T^*_R\right)\Omega,\label{Omega_relacion2}\\
D_m\Omega&=&\frac{1}{2}\left(\frak{g}^{mR}+Q_{m,R}\right)\Omega.\label{Omega_relacion3}
\end{eqnarray}

Again, if we start with the point of view that we want to construct a conformal metric from a system of PDEs, these relations must be satisfied to assure the existence of conformal metrics. Note that these equations do not determine completely the conformal factor, but they are necessary for the conformal equivalence between the ($n-2$)-parameter family of metrics. 

In this way, we have proved that, in particular, all $n$-dimensional spacetime can be considered as originated as the solution space of a  $n(n-3)/2$ PDEs. Note that in  $4$ dimensions, the system is relatively simpler that its nearby $5$-dimensional system, where one must consider five PDEs.

In a similar way as in 4-dimensions, one could construct the Cartan normal conformal connections associated to these PDEs, and from this reduce the system to one which is compatible with Einstein spaces\cite{etal,KNN,GKNP,Nur}. However, although in principle this generalization is direct, in practice the algebraical manipulation of these equations appears as a tedious work , even using algebraic manipulators as Maple or Mathematica.

\section{The Einstein's equations}
If we want to fix completely the conformal metric an extra condition must be imposed.
As we wish to extend NSF to higher dimensions, we will impose the Einstein equations to the system.

As in four dimensions, the vaccum Einstein equations can be obtained by requiring $R^{00}=0$. This equation determine the conformal factor, which is necessary to convert the different conformal metrics in physical ones.
We now adopt a global point of view toward geometry on an
$n$-dimensional manifold. Instead of a metric
$g^{ab}(x^{a})$ on ${\cal M}$, as the fundamental variable we
consider as the basic variables a family of surfaces on ${\cal M}$
given by $u = constant = Z(x^a,s,s^*,\gamma^i)$ or preferably its second
derivatives with respect to $(s,s^*,\gamma^i)$. From this new
point of view, these surfaces are basic and the metric is a derived
concept. Now we will find the conditions on $u = Z(x^a,s,s^*,\gamma^i)$
or more accurately on the second order system such that the
$n$-dimensional metric, be a solution to
the Einstein equations.

We start with the Einstein equations in $n$ dimensions, which are
given by (see for example \cite{Myers,Wh}).

\begin{equation}
R_{ab}= 8\pi G_{(n)} \left(T_{ab}-\frac{1}{n-2}g_{ab}T\right),
\end{equation}
with $G_{(n)}$ the gravitational constant in higher dimensions.

The Ricci tensor is given by
\begin{equation}
R_{ab} = \frac{1}{\sqrt{-g}} \frac{\partial}{\partial x^c}
(\Gamma^c_{ab} \sqrt{-g}) - \frac{\partial^2}{\partial x^a
\partial x^b} \ln \sqrt{-g}- \Gamma^c_{ad}\Gamma^d_{bc},
\label{Rab}
\end{equation}
with $g = \det(g_{ab})$ and
\begin{equation}
\Gamma^c_{ab} = \frac{1}{2} g^{cd}\left(\frac{\partial
g_{da}}{\partial x^b} + \frac{\partial g_{db}}{\partial x^a} -
\frac{\partial g_{ab}}{\partial x^d} \right),  \label{CS}
\end{equation}
the Christoffel symbols.

As happens in $4$-dim, the Einstein equations are given by

\begin{equation}
R^{ab}Z,_{a}Z,_{b}=8\pi G_{(n)} \left(T^{ab}Z,_{a}Z,_{b}\right).
\label{EE}
\end{equation}

That is, to obtain the Einstein equation in this case, we need to
compute $R^{00}\equiv R^{ab}Z,_{a}Z,_{b}$, which is \textit{one}
of the components of $R^{AB}\equiv R^{ab}\theta _{a}^{A}\theta
_{b}^{B}$. From the form of the metric, eq.(\ref{gAB}) we have that
$R^{00}=\Omega^4R_{RR}$.

Before the computation of the equation for the conformal factor $\Omega^2$, we will take the following definitions:\\

1) The contravariants components of the metric $g^{AB}$ will be written $g^{AB}=\Omega^2\frak{g}^{AB}$; in a similar way, we will write
$g_{AB}=\Omega^{-2}\frak{g}_{AB}$.\\

2) Latin indices $A,,B,...,K$, etc. belong to the set
$\{0,+,-,m,R\}$, while the indices
$\{+,-,m\}$ will be denoted with Greek letters:$\{\alpha,\beta,\kappa\}$.\\

3) The determinant of $g^{AB}$ and
$g_{AB}$ will be decomposed as:
\begin{eqnarray}
det(g^{AB})&=&\Omega^{2n}q,\\
det(g_{AB})&=&\Omega^{-2n}\frac{1}{q}=\Omega^{-2n}\Delta.
\end{eqnarray}

4) The derivative $\frac{\partial}{\partial R}$ will be denoted
$\frak{D}$.\\\\

Let us compute now $R_{RR}$:
\begin{equation}
R_{RR}=\frac{1}{\sqrt{-g}} \frac{\partial}{\partial x^C}
(\Gamma^C_{RR} \sqrt{-g}) - \frak{D}^2 \left [\ln \sqrt{-g}\right ]-
\Gamma^C_{RD}\Gamma^D_{RD}. \label{Rrr}
\end{equation}
Note that the only not vanishing term of  $\Gamma^C_{RR}$, is 
$\Gamma^R_{RR}=-2\frac{\frak{D}\Omega}{\Omega}$. Therefore the first term of $R_{RR}$ reads:
\begin{equation}
\frac{1}{\sqrt{-g}} \frac{\partial}{\partial x^C} (\Gamma^C_{RR}
\sqrt{-g})=-2\frac{\frak{D}^2\Omega}{\Omega}-\frac{\frak{D}\Omega}{\Omega}\frac{\frak{D}\Delta}{\Delta}+2(n+1)\frac{(\frak{D}\Omega)^2}{\Omega^2}.
\end{equation}
For the second term, we get:
\begin{equation}
\frak{D}^2 \left [\ln \sqrt{-g}\right
]=-n\frac{\frak{D}^2\Omega}{\Omega}+n\frac{(\frak{D}\Omega)^2}{\Omega^2}+\frac{\frak{D}^2\Delta}{2\Delta}-\frac{(\frak{D}\Delta)^2}{2\Delta^2}.
\end{equation}
Finally, let us compute the remaining terms of $R_{RR},$ i.e.:
\begin{equation}
\Gamma^C_{RD}\Gamma^D_{RD}=\Gamma^0_{RD}\Gamma^D_{R0}+\Gamma^\alpha_{RD}\Gamma^D_{R\alpha}+\Gamma^R_{RD}\Gamma^D_{RR}.
\end{equation}
Due to the fact that
$\Gamma^0_{RD}=\Gamma^\alpha_{R0}=0$ and that the only not vanishing term of $\Gamma^D_{RR}$ is $\Gamma^R_{RR}$, such an expression is reduced to
\begin{equation}
\Gamma^C_{RD}\Gamma^D_{RC}=\Gamma^\alpha_{R\beta}\Gamma^\beta_{R\alpha}+\left
(\Gamma^R_{RR}\right )^2.
\end{equation}
Using the decomposition $g^{AB}=\Omega^2\frak{g}^{AB}$,
$g_{AB}=\Omega^{-2}\frak{g}_{AB}$, one finds:
\begin{equation}
\Gamma^\alpha_{R\beta}=-\frac{\frak{D}\Omega}{\Omega}\delta^\alpha_\beta-\frac{1}{2}\frak{D}[\frak{g}^{\alpha\kappa}]{\frak{g}_{\beta\kappa}}.
\end{equation}
Therefore,
\begin{equation}
\Gamma^\alpha_{R\beta}\Gamma^\beta_{R\alpha}=(n-2)\frac{(\frak{D}\Omega)^2}{\Omega^2}+\frac{\frak{D}\Omega}{\Omega}\frak{D}[\frak{g}^{\alpha\beta}]\frak{g}_{\alpha\beta}
+\frac{1}{4}\frak{D}[\frak{g}^{\alpha\epsilon}]\frak{D}[\frak{g}^{\beta\kappa}]\frak{g}_{\beta\epsilon}\frak{g}_{\alpha\kappa}.
\end{equation}
Adding all these terms we obtain
\begin{eqnarray}
R_{RR}&=&(n-2)\frac{\frak{D}^2\Omega}{\Omega}-\frac{\frak{D}\Omega}{\Omega}\left
(\frac{\frak{D}\Delta}{\Delta}+\frak{D}[\frak{g}^{\alpha\beta}]\frak{g}_{\alpha\beta}\right
)\nonumber\\&&-\frac{\frak{D}^2\Delta}{2\Delta}+\frac{(\frak{D}\Delta)^2}{2\Delta^2}+\frac{1}{4}\frak{D}[\frak{g}^{\alpha\epsilon}]\frak{D}[\frak{g}^{\beta\kappa}]\frak{g}_{\beta\epsilon}\frak{g}_{\alpha\kappa};\label{5rr}
\end{eqnarray}
but the second term is vanishing, since as is well known from properties of determinants (see for example \cite{Poisson}),
\begin{equation}
\frac{\frak{D}\Delta}{\Delta}+\frak{D}[\frak{g}^{\alpha\beta}]\frak{g}_{\alpha\beta}=0.
\end{equation}
Finally, from (\ref{5rr}), we get the equation that determine $\Omega$, \textit{the Einstein equations},
where in order to compare with the $4$-dim case discussed in the literature, we have replaced $\Delta$ by $1/q$,
\begin{equation}
\frak{D}^2\Omega=\frac{8\pi G_{(n)}}{n-2}
T^{00}\Omega^{-3}+\frac{1}{n-2}\left [\frac{1}{2}\frac{(\frak{D}q)^2}{q^2}-\frac{\frak{D}^2q}{2q}+\frac{1}{4}\frak{D}[\frak{g}^{\alpha\epsilon}]\frak{D}[\frak{g}^{\beta\kappa}]\frak{g}_{\beta\epsilon}\frak{g}_{\alpha\kappa}\right
]\Omega.\label{NSFomegandim}
\end{equation}
This equation in addition to the metricity conditions and the other three relations eqs.(\ref{Omega_relacion1})-(\ref{Omega_relacion3}) that $\Omega$ must satisfy, constitute a system equivalent to the ten Einstein equations for the metric $g_{ab}$.

As a particular case, we cite the well-known case of NSF in 4 dimensions, where one has a system of two PDEs, namely:
\begin{eqnarray}
\partial_{s s}Z & = & S_{}(u,w^+,w^-,R,s,s^*),\label{ss} \\
\partial_{s^* s^*}Z & = & S^*_{}(u,w^+,w^-,R,s,s^*)\label{ss*}. 
\end{eqnarray}

The contravariants components of the metric are
\begin{equation}
g^{AB}=\Omega^2\frak{g}^{AB}=\Omega^2\left(%
\begin{array}{cccc}
  0 & 0 & 0 & 1 \\
  0 & a & -1 & b \\
  0 & -1 & a^* & b^* \\
  1 & b & b^* & c \\
\end{array}%
 \right),
\end{equation}
with $a=-S_{R}$ y $a^*=-S^*_{R}$. (The expressions for $b,b^*$ and $c$ are not necessary to write the Einstein equations).\\
From this expression for the metric we have that
\begin{equation}
\frac{1}{4}\frak{D}[\frak{g}^{\alpha\epsilon}]\frak{D}[\frak{g}^{\beta\kappa}]\frak{g}_{\beta\epsilon}\frak{g}_{\alpha\kappa}
=\frac{1}{4}\frac{(\frak{D}q)^2}{q^2}+\frac{1}{2}\frac{\frak{D}a\frak{D}a^*}{q},
\end{equation}
with $q=1-S_RS^*_R$.

In this way by replacing in eq.(\ref{NSFomegandim}) we obtain 
\begin{equation}
\frak{D}^2\Omega=4\pi G
T^{00}\Omega^{-3}+\left [\frac{3}{8}\frac{(\frak{D}q)^2}{q^2}-\frac{1}{4}\frac{\frak{D}^2q}{q}+\frac{1}{4}\frac{\frak{D}^2[S_{}]\frak{D}^2[S^*_{}]}{q}\right
]\Omega,
\end{equation}
and therefore, we recover the well-known equation for the conformal factor obtained first (with an error of a factor $-1$)  in~\cite{FCN3} (see the correction given in ~\cite{Erratum}, in equation $(5)$).

\section{NSF of GR in 5 dimensions}
As an explicit case, we will consider now the NSF of GR in five dimensions.  The motivation is twofold, on one hand, the Einstein equations in five dimensions can be used as an attempt of a geometrical unification between the Maxwell electrodynamics (more precisely null electromagnetic fields) and  gravitational fields in the manner of \textit{Kaluza-Klein}. Although it is not so clear if this program can be studied in term of NSF, it would be very interesting to be able to put both fields in terms of the function $Z$. 
On the other hand, this is the most simple example of the formalism that one can study in dimensions higher than four. We also present some simple known metrics expressed in this formalism. 
\subsection{The equations}
The null surfaces can be associated with general solutions from the following system of five PDEs, 

\begin{eqnarray}
\partial_{ss}Z&=&S(Z,\partial_s Z,\partial_{s^*}Z,\partial_{\gamma}Z,\partial_{ss^*}Z,s,s^*,\gamma),\label{pde51}\\
\partial_{s^*s^*}Z&=&S^*(Z,\partial_s Z,\partial_{s^*}Z,\partial_{\gamma}Z,\partial_{ss^*}Z,s,s^*,\gamma),\\
\partial_{s\gamma}Z&=&\Phi(Z,\partial_s Z,\partial_{s^*}Z,\partial_{\gamma}Z,\partial_{ss^*}Z,s,s^*,\gamma),\\
\partial_{s^*\gamma}Z&=&\Phi^*(Z,\partial_s Z,\partial_{s^*}Z,\partial_{\gamma}Z,\partial_{ss^*}Z,s,s^*,\gamma),\\
\partial_{\gamma\gamma}Z&=&\Upsilon(Z,\partial_s Z,\partial_{s^*}Z,\partial_{\gamma}Z,\partial_{ss^*}Z,s,s^*,\gamma).\label{pde52}
\end{eqnarray}
These functions must satisfy the W\"{u}nschmann conditions (\ref{W1})-(\ref{W2}),
\begin{eqnarray}
\frak{m}&=&(D_sS)_{R}-3\left(S_{w^+}S_R+S_{w^-}+S_{w^1}\Phi_R-S_R\frak{g}^{+R}\right)=0,\label{w51}\\
\frak{m}^*&=&(D_{s^*}S^*)_{R}-3\left(S^*_{w^-}S^*_R+S^*_{w^+}+S^*_{w^1}\Phi^*_R-S^*_R\frak{g}^{-R}\right)=0,\\
\frak{m}_{111}&=&(D_1\Upsilon)_{R}-3\left(\Upsilon_{w^+}\Phi_{R}+\Upsilon_{w^-}\Phi^*_{R}+\Upsilon_{w^1}\Upsilon_{R}-\Upsilon_{R}\frak{g}^{1R}\right)=0,\\
\frak{m}_{1}&=&(D_1S)_{R}-\left(S_{w^-}\Phi^*_{R}+S_{w^+}\Phi_{R}+S_{w^1}\Upsilon_{R}-S_{R}\frak{g}^{1R}\right)\nonumber\\&&-2\left(\Phi_{w^-}+\Phi_{w^+}S_{R}+\Phi_{w^1}\Phi_{R}-\Phi_{R}\frak{g}^{+R}\right)=0,\\
\frak{m}^*_{1}&=&(D_1S^*)_{R}-\left(S^*_{w^+}\Phi_{R}+S^*_{w^-}\Phi^*_{R}+S^*_{w^1}\Upsilon_{R}-S^*_{R}\frak{g}^{1R}\right)\nonumber\\&&-2\left(\Phi^*_{w^+}+\Phi^*_{w^-}S^*_{R}+\Phi^*_{w^1}\Phi^*_{R}-\Phi^*_{R}\frak{g}^{-R}\right)=0,\\
\frak{m}_{11}&=&(D_s\Upsilon)_{R}-\left(\Upsilon_{w^+}S_{R}+\Upsilon_{w^-}+\Upsilon_{w^1}\Phi_{R}\right)\nonumber\\
&&-2\left(\Phi_{w^+}\Phi_{R}+\Phi_{w^-}\Phi^*_{R}+\Phi_{w^1}\Upsilon_{R}\right)+\Upsilon_{R}\frak{g}^{+R}+2\Phi_{R}\frak{g}^{1R}=0,\\
\frak{m}^*_{11}&=&(D_{s^*}\Upsilon)_{R}-\left(\Upsilon_{w^-}S^*_{R}+\Upsilon_{w^+}+\Upsilon_{w^1}\Phi^*_{R}\right)\nonumber\\
&&-2\left(\Phi_{w^-}\Phi_{R}+\Phi^*_{w^+}\Phi_{R}+\Phi^*_{w^1}\Upsilon_{R}\right)+\Upsilon_{R}\frak{g}^{-R}+2\Phi^*_{R}\frak{g}^{1R}=0.\label{w52}
\end{eqnarray}
In these relations, $\frak{g}^{+R},\frak{g}^{-R}$ and $\frak{g}^{1R}$ are some of the components of the conformal metric
\begin{eqnarray}
\frak{g}^{+R}&=&-\frac{1}{4-S_{R}S^*_{R}}
\left[2(T_R-S_{w^+}-S_{w^-}S^*_{R}
-S_{w^1}\Phi^*_{R})\right.\nonumber\\
&&\left. -S_{R}(T^*_R-S^*_{w^-}-S^*_{w^+}S_{R}
-S^*_{w^1}\Phi_{R})\right],\label{g+R} \\
\frak{g}^{-R}&=&-\frac{1}{4-S_{R}S^*_{R}}
[2(T^*_R-S^*_{w^-}-S^*_{w^+}S_{R}
-S^*_{w^1}\Phi_{R})\nonumber\\
&&-S^*_{R}(T_R-S_{w^+}-S_{w^-}S^*_{R}
-S_{w^1}\Phi^*_{R})],\label{g-R} \\
\frak{g}^{1R}&=&-\Phi^*_{R}\frak{g}^{+R}- \Phi_{R}\frak{g}^{-R}-(Q_{1R}
-\Phi_{w^+}-\Phi_{w^-}S^*_{R}\nonumber\\
&&-\Phi_{w^1}\Phi^*_{R}-\Phi^*_{w^+}S_{R}
-\Phi^*_{w^-}-\Phi^*_{w^1}\Phi_{R}).\label{g1R}
\end{eqnarray}
\newline

\textbf{Remark 3:} \textit{ Note also, that using the commutator relations (valid in $n$-dimensions),
\begin{eqnarray}
\left[\partial_y,D_{s}\right]&=&\delta_{w^+,y}\partial_u+S_{.y}\partial_{w^+}+\delta_{R,y}\partial_{w^-}+\Phi_{k,y}\partial_{w^k}+T_y\partial_R,\\
\left[\partial_y,D_{s^*}\right]&=&\delta_{w^-,y}\partial_u+\delta_{R,y}\partial_{w^+}+S^*_{.y}\partial_{w^-}+\Phi^*_{k,y}\partial_{w^k}+T^*_y\partial_R,\\
\left[\partial_y,D_{m}\right]&=&\delta_{w^m,y}\partial_u+\Phi_{m,y}\partial_{w^+}+\Phi^*_{m.y}\partial_{w^-}+\Upsilon_{km,y}\partial_{w^k}+Q_{m,y}\partial_R,
\end{eqnarray}
with $y\in\{u,w^+,w^-,w^m,R\}$ and $\delta_{y',y}$ the Kronecker symbol, we have
\begin{eqnarray}
(D_{s}S)_R&=&D_s(S_R)+S_{R}S_{w^+}+S_{w^-}+\Phi_{R}S_{w^1}+T_RS_R,\\
(D_{s^*}S^*)_R&=&D_{s^*}(S^*_R)+S^*_{R}S^*_{w^-}+S^*_{w^+}+\Phi^*_{R}S^*_{w^1}+T^*_RS^*_R,\\
(D_{1}S)_R&=&D_1(S_R)+\Phi_{R}S_{w^+}+\Phi^*_{R}S_{w^-}+\Upsilon_{R}S_{w^1}+Q_{1,R}S_R,\\
(D_{1}S^*)_R&=&D_1(S^*_R)+\Phi^*_{R}S^*_{w^-}+\Phi_{R}S^*_{w^+}+\Upsilon_{R}S^*_{w^1}+Q_{1,R}S^*_R,\\
(D_{s}\Upsilon)_R&=&D_s(\Upsilon_R)+S_{R}\Upsilon_{w^+}+\Upsilon_{w^-}+\Phi_{R}\Upsilon_{w^1}+T_R\Upsilon_R,\\
(D_{s^*}\Upsilon)_R&=&D_{s^*}(\Upsilon_R)+S^*_{R}\Upsilon_{w^-}+\Upsilon_{w^+}+\Phi^*_{R}\Upsilon_{w^1}+T^*_R\Upsilon_R,\\
(D_{1}\Upsilon)_R&=&D_1(\Upsilon_R)+\Phi_{R}\Upsilon_{w^+}+\Phi^*_{R}\Upsilon_{w^-}+\Upsilon_{R}\Upsilon_{w^1}+Q_{1,R}\Upsilon_R.
\end{eqnarray}}\newline

The other components of the conformal metric are
\begin{eqnarray}
\frak{g}^{00}&=&\frak{g}^{0+}=\frak{g}^{0-}=\frak{g}^{01}=0,\label{g01}\\
\frak{g}^{0R}&=&-\frak{g}^{+-}=1,\\
\frak{g}^{++}&=&-S_R,\;\;\frak{g}^{--}=-S^*_R,\\
\frak{g}^{+1}&=&-\Phi_R,\;\;\frak{g}^{-1}=-\Phi^*_R,\\
\frak{g}^{11}&=&-\Upsilon_R,\label{g11}
\end{eqnarray}
and
\begin{eqnarray}
\frak{g}^{RR}&=&-\frac{1}{2+S_RS^*_R}\left\{S^*_{w^1}S_R \frak{g}^{1R}+\left[U_R-2T_{w^+}
-2T_{w^-}S^*_{R}-2T_{w^1}\Phi^*_{R}-2T^*_{w^+}S_{R}\right.\right.\nonumber\\
&&\left.\left.-2T^*_{w^-}-2T^*_{w^1}\Phi_{R}+S^*_{u}S_{R}+S^*_{R}S_{u}-S^*_{w^+}\left(S_{w^+}S_{R}+S_{w^-}+S_{w^1}\Phi_{R}\right)\right.\right.\nonumber\\
&&\left.\left.-S^*_{w^-}\left(S_{w^+}+S_{w^-}S^*_{R}+S_{w^1}\Phi^*_{R}\right)-S^*_{w^1}\left(S_{w^+}\Phi_{R}+S_{w^-}\Phi^*_{R}+S_{w^1}\Upsilon_{R}\right)\right]\frak{g}^{0R}\right.\nonumber\\
&&\left.+\left(2T^*_R+S^*_{w^+}S_R+S^*_RS_{w^+}\right)\frak{g}^{+R}+\left(2T_R+S_{w^-}S^*_R+S_RS^*_{w^-}\right)\frak{g}^{-R}\right\}.\label{gRR}
\end{eqnarray}

Finally the conformal factor must satisfy the relations
\begin{eqnarray}
D_s\Omega&=&\frac{1}{2}\left(\frak{g}^{+R}+T_R\right)\Omega,\label{Omega_relacion1}\\
D_{s^{*}}\Omega&=&\frac{1}{2}\left(\frak{g}^{-R}+T^*_R\right)\Omega,\label{Omega_relacion2}\\
D_1\Omega&=&\frac{1}{2}\left(\frak{g}^{1R}+Q_{m,R}\right)\Omega,\label{Omega_relacion3}
\end{eqnarray}
and the Einstein equation (\ref{NSFomegandim}),
\begin{eqnarray}
\frak{D}^2\Omega&=&\frac{8\pi G_{(5)}}{3}T^{00}\Omega^{-3}+\frac{1}{6}\left\{\frac{3}{2}\frac{(\frak{D}q)^2}{q^2}-\frac{\frak{D}^2q}{q}+\frac{1}{q}\left[2\left(\Phi_R\frak{D}\Phi_R\frak{D}S^*_R+\Phi^*_R\frak{D}\Phi^*_R\frak{D}S_R-\frak{D}\Phi_R\frak{D}\Phi^*_R\right)\right.\right.\nonumber\\
&&\left.\left.-S_R\frak{D}\Upsilon_R\frak{D}S^*_R-S^*_R\frak{D}\Upsilon_R\frak{D}S_R-\Upsilon_R\frak{D}S_R\frak{D}S^*_R+S_R(\frak{D}\Phi_R)^2+S^*_R(\frak{D}\Phi^*_R)^2\right]\right\}\Omega,\label{eq:einstconf}
\end{eqnarray}
with
\begin{equation}
q=2\Phi_R\Phi^*_R-S_R\Phi^{*2}_R-S^*_R\Phi^{2}_R-\Upsilon_R(1-S_RS^*_R).
\end{equation}

\subsection{A simple example}

Let there be the following family of null surfaces of the 5-dim Minkowski spacetime,
\begin{equation}
u=Z(x^a,\zeta,\bar\zeta,\gamma)=x^al_a(\zeta,\bar\zeta,\gamma),\label{repre}
\end{equation}
with $x^a=\{t,x,y,z,v\}$ Minkowskian coordinates, and $l_a$  the covariant components of the null vector $l^a$ given by
\begin{equation}
l^a(\zeta,\bar\zeta,\gamma)=\frac{1}{\sqrt{2}(1+\zeta\bar\zeta)}((1+\zeta\bar\zeta),(\zeta+\bar\zeta)\sin\gamma,i(\bar\zeta-\zeta)\sin\gamma,(\zeta\bar\zeta-1)\sin\gamma,(1+\zeta\bar\zeta)\cos\gamma),
\end{equation}
with $\{\zeta,\bar\zeta,\gamma\}$ the coordinates on a $3$-dimensional sphere $S^3$, and $\bar\zeta$ the complex conjugate of $\zeta$.\\  

We will make the following identification between parameters: $s\equiv \zeta, s^*\equiv\bar\zeta$, and $\gamma^1\equiv\gamma$.\\

Therefore, from these relations we can construct the scalars $\theta ^{A}$,
\begin{eqnarray}
u&=&\frac{\sqrt{2}}{2}\left\{t-\frac{\sin\gamma}{1+\zeta\bar\zeta}\left[(\zeta+\bar\zeta)x+i(\bar\zeta-\zeta)y-(1-\zeta\bar\zeta)z\right]-\cos\gamma\, v \right\},\\
w^+&=&\frac{\sqrt{2}}{2}\frac{\sin\gamma}{(1+\zeta\bar\zeta)^2}\left[(\bar\zeta^{2}-1)x+i(\bar\zeta^{2}+1)y-2\bar\zeta z\right],\\
w^-&=&\frac{\sqrt{2}}{2}\frac{\sin\gamma}{(1+\zeta\bar\zeta)^2}\left[(\zeta^2-1)x-i(\zeta^2+1)y-2\zeta z\right],\\
w^1&=&\frac{\sqrt{2}}{2}\left\{-\frac{\cos\gamma}{1+\zeta\bar\zeta}\left[(\zeta+\bar\zeta)x+i(\bar\zeta-\zeta)y-(1-\zeta\bar\zeta)z\right]+\sin\gamma\, v \right\},\\
R&=&\sqrt{2}\frac{\sin\gamma}{(1+\zeta\bar\zeta)^3}\left[(\zeta+\bar\zeta)x+i(\bar\zeta-\zeta)y-(1-\zeta\bar\zeta)z\right].
\end{eqnarray}
Solving these equations for the $x^a$'s we obtain
\begin{eqnarray}
t&=&\sqrt{2}\left[u+\cot\gamma\, w^1+\frac{(1+\zeta\bar\zeta)^2}{2\sin^2\gamma}R \right],\label{eq:trans1}\\
x&=&\frac{\sqrt{2}}{2\sin\gamma}\left[(\bar\zeta-1) w^- +(\zeta^2-1)w^+ +(\zeta+\bar\zeta)(1+\zeta\bar\zeta)R \right],\\
y&=&\frac{\sqrt{2}i}{2\sin\gamma}\left[(\bar\zeta+1) w^- -(\zeta^2+1)w^+ +(\zeta-\bar\zeta)(1+\zeta\bar\zeta)R \right],\\
z&=&-\frac{\sqrt{2}}{2\sin\gamma}\left[2\bar\zeta w^- +2\zeta w^+ +(1-\zeta^2\bar\zeta^{2})R \right],\\
v&=&\frac{\sqrt{2}}{2\sin\gamma}\left[ w^1 +\cos\gamma(1+\zeta\bar\zeta)^2R \right]\label{eq:trans2}.
\end{eqnarray}

Remember that these equations can be interpreted as a three-parameter family of coordinate transformations between $x^a$ and $\{u,w^+,w^-,w^1,R\}$.
In particular, for fixed values of $\zeta,\bar\zeta,$ and $\gamma$ we can write the Minkowskian metric in the new coordinates $\{u,w^+,w^-,w^1,R\}$, 
\begin{eqnarray}
ds^2&=&\eta_{ab}dx^adx^b=\frac{\partial x^a}{\partial\theta^A}\frac{\partial x^b}{\partial\theta^B}\eta_{ab}d\theta^Ad\theta^B\nonumber\\
&=&2du^2+2\frac{(1+\zeta\bar\zeta)^2}{\sin^2\gamma}dudR+2\cot\gamma\,dw^1du-2\frac{(1+\zeta\bar\zeta)^2}{\sin^2\gamma}\,dw^+dw^--2(dw^1)^2.\label{Minkowski}
\end{eqnarray}

On the other hand, from eq.(\ref{tildef1})-(\ref{tildef2}), and the transformations (\ref{eq:trans1})-(\ref{eq:trans2}), we see that
\begin{eqnarray}
S&=&-\frac{2\bar\zeta}{1+\zeta\bar\zeta}w^+,\label{S1}\\
S^*&=&-\frac{2\zeta}{1+\zeta\bar\zeta}w^-,\label{S2}\\
\Phi &=&\cot\gamma\, w^+,\\
\Phi^* &=&\cot\gamma\,w^-,\\
\Upsilon &=&\frac{1}{2\sin^2\gamma}\left[(1+\zeta\bar\zeta)^2R+2\sin\gamma\cos\gamma\,w^1\right].
\end{eqnarray}
In a similar way, we obtain
\begin{eqnarray}
T&=&D_{\bar\zeta}S=-\frac{2}{(1+\zeta\bar\zeta)^2}\left[w^+ +\bar\zeta(1+\zeta\bar\zeta)R\right],\\
T&=&D_{\zeta}S^*=-\frac{2}{(1+\zeta\bar\zeta)^2}\left[w^- +\zeta(1+\zeta\bar\zeta)R\right],\\
Q_1&=&D_{\bar\zeta}\Phi=\cot\gamma\,R.
\end{eqnarray}

Now instead of starting with the flat metric (\ref{Minkowski}), we want to consider the system of PDEs (\ref{pde51})-(\ref{pde52}) and reconstruct from it the (conformal) metric.

It is an easy task to shown that, as it should be, the functions ($S,S^*,\Phi,\Phi^*,\Upsilon$) satisfy the W\"{u}nshmann conditions (\ref{w51})-(\ref{w52}).

Finally from the expressions eqs.(\ref{g+R})-(\ref{g1R}) and (\ref{g01})-(\ref{gRR}) for the contravariant components of the (conformal) metric in terms of ($S,S^*,\Phi,\Phi^*,\Upsilon$) we get that the only non vanishing components are
\begin{eqnarray}
g^{0R}&=&\Omega^2,\\
g^{+-}&=&-\Omega^2,\\
g^{RR}&=&-\frac{2}{(1+\zeta\bar\zeta)^2}\Omega^2,\\
g^{1R}&=&\cot\gamma\,\Omega^2,\\
g^{11}&=&-\frac{(1+\zeta\bar\zeta)^2}{\sin^2\gamma}\Omega^2;
\end{eqnarray}
or in its covariant version
\begin{equation}\label{gAB4}
g_{AB}=\Omega^{-2}
\left(%
\begin{array}{ccccc}
  \frac{2\sin^2\gamma}{(1+\zeta\bar\zeta)^2} & 1 & 0  & 0 & \frac{2\sin\gamma\cos\gamma}{(1+\zeta\bar\zeta)^2} \\
  1 & 0 & 0  & 0 & 0 \\
  0 & 0 & 0 &  -1 & 0 \\
  0 & 0 & -1 &  0 & 0 \\
  \frac{2\sin\gamma\cos\gamma}{(1+\zeta\bar\zeta)^2}& 0 & 0&  0 & -\frac{2\sin^2\gamma}{(1+\zeta\bar\zeta)^2}\\
\end{array}%
\right) .
\end{equation}
On the other hand the conformal factor must satisfy eqs.(\ref{Omega_relacion1})-(\ref{Omega_relacion3}),
\begin{eqnarray}
D_\zeta\Omega &=&\frac{T_R}{2}\Omega=-\frac{\bar\zeta}{1+\zeta\bar\zeta}\Omega,\label{om1}\\
D_{\bar\zeta}\Omega &=&\frac{T^*_R}{2}\Omega=-\frac{\zeta}{1+\zeta\bar\zeta}\Omega,\label{om2}\\
D_\gamma\Omega &=&\frac{1}{2}\left(\frak{g}^{1R}+Q_{1,R}\right)\Omega=\cot\gamma\,\Omega,\label{om3}
\end{eqnarray} 
and the Einstein's equation (\ref{eq:einstconf}), (with $T^{00}=0$)
\begin{equation}
\frak{D}^2\Omega=0.\label{om4}
\end{equation}
This system  admits as a particular solution 
\begin{equation}
\Omega_\eta=\frac{\sin\gamma}{1+\zeta\bar\zeta}.
\end{equation}
Therefore, it is a simple task to show that this choice makes the metric $g_{AB}$, the Minkowski metric. eq.(\ref{Minkowski}).

However the system (\ref{om1})-(\ref{om4}) admits another conformal flat solutions, as for example the de Sitter solution,  
\begin{eqnarray}
\Omega&=&\Omega_\eta+\Lambda\eta_{ab}x^ax^b\\
&=&\frac{\sin\gamma}{1+\zeta\bar\zeta}+2\Lambda\left[u^2-(w^1)^2+2\cot\gamma w^1u+\frac{(1+\zeta\bar\zeta)^2}{\sin^2\gamma}(uR-w^+w^-)\right],
\end{eqnarray}
with $\Lambda=\text{constant}.$\\\\

\textbf{Remark 4:} \textit{In the four-dimensional case, the system is given by the pair of PDEs eqs.(\ref{ss})-(\ref{ss*}), 
with the two functions $S$ and $S^*$ as in eqs.(\ref{S1})-(\ref{S2}). 
On the other hand, in the literature~\cite{FCN3} one can find that the conformal flat metrics can be obtained from the most simple system of equations 
\begin{eqnarray}
\eth^2 Z&=&0,\\
\bar\eth^2 Z&=&0,
\end{eqnarray}
where $\eth$ is the \textit{edth} operator that acts on a function $\eta$ of spin-weight $s$ as $\eth\eta=2P^{1-s}\partial_\zeta(P^s\eta),$ and $P=\frac{1}{2}(1+\zeta\bar\zeta).$  
There is not incompatibility between these two systems, in fact acting on a function $Z$ ($s=0$), we have
\begin{equation}
\eth^2 Z=4\partial_\zeta(P^2\partial_\zeta Z)=4P^2\partial_{\zeta\zeta}Z+8P\partial_\zeta P\partial_\zeta Z;
\end{equation}
and therefore from $\eth^2 Z=0,$ we get 
\begin{equation}\partial_{\zeta\zeta}Z=-2\partial_{\zeta}(\ln P)\partial_\zeta Z=-\frac{2\bar\zeta}{1+\zeta\bar\zeta}\partial_\zeta Z;
\end{equation} which coincides with our $S$. 
It would be desirable to reformulate NSF in higher dimensions in terms of covariants operators on the sphere $S^{n-2}$.}

\section{Final comments}
We have shown that the null surface formulation can be extended to higher dimensions. In particular all conformal $n$-dim metric can be constructed from a particular class of $n(n-3)/2$ PDEs. In order to assure the existence of the metric, this class must satisfy a set of metricity conditions. However this system of PDEs becomes more involved when the considered dimension increases, and it is no so easy to write explicit conditions in order the metric to be Lorentzian. This in an important caveat. Yet in this case, it is notable that all the information contained in the $n(n+1)/2$ metric components (local point of view), can be globally codified in only two functions $Z$ and $\Omega$ dependent of $n-2$ parameters (global point of view). 
     
On the other hand, it would be expected that the W\"{u}nschmann conditions have a similar geometrical meaning as in $3$ and $4$ dimensions. In the last two cases, one can show that these conditions can be understood as the requirement of a vanishing torsion tensor of a canonical connection defined on the solution space associated to the differential equations.  In fact, if there exists such connection, then the Lie derivative of the metric can be put in terms of the torsion components, and if the torsion vanishes, the Lie derivative of the metric is proportional to the metric, i.e., they are all in the same conformal class. Although in higher dimensions the equations are more complicated, the program should in principle be possible. These results will be presented elsewhere. 

Finally, as was mentioned before, it should be desirable to express this formalism in term of covariants operators, and from then find a geometrical meaning to the functions $(S,S^*,\Phi_m,\Phi^*_m,\Upsilon_{nm})$.  
 \section*{Acknowledgments}

We acknowledge support from CONICET and SeCyT-UNC.

\section*{Appendix A}
Here we would like to give some examples of families of null surfaces of $n$-dimensional Lorentzian flat spacetimes,  such that eq.(\ref{eq:inverse}) is satisfied.  
Let us begin with the following $(n-2)$-parametric family of functions 
\begin{equation}
u=Z(x^a,s_1,s_2,\cdots,s_{n-2})=x_0-\frac{2s_1}{1+r^2}x_1-\frac{2s_2}{1+r^2}x_2-\cdots-\frac{2s_{n-2}}{1+r^2}x_{n-2}-\frac{r^2-1}{1+r^2}x_{n-1}.\label{eq:zapend}
\end{equation}
with $r^2=s_1^2+s_2^2+\cdots+s_{n-2}^2$ and $\{s_1,s_2,\cdots,s_{n-2}\}\in \mathbb{R}^{n-2}$. The parameters $s_i$ can be thought as stereographic coordinates of the sphere $S^{n-2}$. 
This family of functions satisfy the eikonal equation
\begin{equation}
(Z_{x_0})^2-(Z_{x_1})^2-\cdots-(Z_{x_{n-1}})^2=0.
\end{equation} 
and geometrically they represent null plane waves $u=x^al_a\left(s_1,\cdots,s_{n-2}\right)$ spanned by the null vectors
\begin{equation}
l^a=\left(1,\frac{2s_1}{1+r^2},\cdots,\frac{2s_{n-2}}{1+r^2},\frac{r^2-1}{1+r^2}\right).
\end{equation}
Now, if we define the scalars $\theta^A,_a$ from  this $Z$ (with any identification between the parameters\\
 $\{s_1,s_2,\cdots,s_{n-2}\}$ and $\{s,s^*,\cdots,\gamma^{m}\}$), we get that $\text{det}\;\theta^A,_a=0$, i.e., they are not independent scalars.
However by doing the following (complex) transformation,
\begin{eqnarray}
s&\equiv & \zeta=s_1+is_2,\\
s^*&\equiv &\bar\zeta=s_1-is_2,\\
\gamma^m&=&s_m,
\end{eqnarray}
we get that the new $\theta^A,_a$ satisfy
\begin{equation} 
\text{det}\;\theta^A,_a= -\left(\frac{-2}{1+r^{*2}}\right)^{n-2}i,
\end{equation}
with $r^{*2}=1+\zeta\bar\zeta+\gamma_1^2+\cdots+\gamma_{n-2}^2$. Therefore eq.(\ref{eq:inverse}) is satisfied (with the exception of the pole $r^*\rightarrow \infty$).

Let us consider now, the family
\begin{equation}
Z=x_0-\tilde s_1x_1-\tilde s_2x_2-\cdots-\tilde s_{n-2}x_{n-2}-\sqrt{1-\tilde r^2}x_{n-1},\label{eq:raiz}
\end{equation}
with $\tilde r^2=\tilde s_1^2+\tilde s_2^2+\cdots+\tilde s_{n-2}^2$ and $s_i \in [-1,1]$. Note, that this family can be obtained from the previous one, eq.(\ref{eq:zapend}) by the transformation,
\begin{eqnarray}
\tilde s_1&=&\frac{2s_1}{1+r^2},\\
\tilde s_2&=&\frac{2s_2}{1+r^2},\\
&\vdots&\nonumber\\
\tilde s_{n-2}&=&\frac{2s_{n-2}}{1+r^2}.
\end{eqnarray}
In this case, by doing the identification $s\equiv s_1$, $s^*\equiv s_2$, and $\gamma^m\equiv s_m$, we obtain
\begin{equation}
\text{det}\;\theta^A\,_{,a}=\frac{(-1)^n s s^*}{(1-\tilde r^2)^{3/2}},
\end{equation}
and therefore the scalars $\theta^A$ are independent in the regions where $ s$ and $s^*$ are non vanishing.

An alternative possibility  was used in the example of $5$-dimensional Minkowski spacetime eq.(\ref{repre}). In that case $\text{det}\;\theta^A\,_{,a}=\frac{4i\sin^4\gamma_1}{(1+\zeta\bar\zeta)^4}$.

Note that the examples given by eq.(\ref{eq:zapend}) and eq.(\ref{eq:raiz}) geometrically represent the same family of null waves, but they give origin to distinct (although  equivalent) partial differential equations. They are related to each other by a fiber-preserving transformation. In fact, in three and four dimensions, it was shown by Frittelli, Kamran and Newman, the equivalence between PDEs with vanishing W\"{u}nschmann invariants under more general tranformations than fiber-preserving known as contact transformations\cite{FKamN,FKamN2,FKamN3}.

\section*{Appendix B}
The explicit expressions for the generalized W\"{u}nschmann conditions (\ref{W1})-(\ref{W2}) are,
\begin{eqnarray}
\frak{m}&=&(D_sS)_{R}-3\left(S_{w^+}S_R+S_{w^-}+S_{w^m}\Phi_{m,R}-S_R\frak{g}^{+R}\right)=0,\label{w51ap}\\
\frak{m}^*&=&(D_{s^*}S^*)_{R}-3\left(S^*_{w^-}S^*_R+S^*_{w^+}+S^*_{w^m}\Phi^*_{m,R}-S^*_R\frak{g}^{-R}\right)=0,\\
\frak{m}_{kmn}&=&(D_k\Upsilon_{mn})_{R}-\Upsilon_{mn,w^+}\Phi_{k,R}-\Upsilon_{mk,w^+}\Phi_{n,R}-\Upsilon_{kn,w^+}\Phi_{m,R}\nonumber\\
&&-\Upsilon_{mn,w^-}\Phi^*_{k,R}-\Upsilon_{mk,w^-}\Phi^*_{n,R}-\Upsilon_{kn,w^-}\Phi^*_{m,R}\nonumber\\&&   -\Upsilon_{mn,w^l}\Upsilon_{lk,R}-\Upsilon_{mk,w^l}\Upsilon_{ln,R}-\Upsilon_{kn,w^l}\Upsilon_{lm,R}\nonumber\\&& 
+\Upsilon_{mn,R}\frak{g}^{kR}+\Upsilon_{mk,R}\frak{g}^{nR}+\Upsilon_{kn,R}\frak{g}^{mR}=0,\\
\frak{m}_{m}&=&(D_mS)_{R}-\left(S_{w^-}\Phi^*_{m,R}+S_{w^+}\Phi_{m,R}+S_{w^n}\Upsilon_{nm,R}-S_{R}\frak{g}^{mR}\right)\nonumber\\&&-2\left(\Phi_{m,w^-}+\Phi_{m,w^+}S_{R}+\Phi_{m,w^n}\Phi_{n,R}-\Phi_{m,R}\frak{g}^{+R}\right)=0,\\
\frak{m^*}_{m}&=&(D_mS^*)_{R}-\left(S_{w^+}\Phi_{m,R}+S^*_{w^-}\Phi^*_{m,R}+S^*_{w^n}\Upsilon_{nm,R}-S^*_{R}\frak{g}^{mR}\right)\nonumber\\&&-2\left(\Phi^*_{m,w^+}+\Phi^*_{m,w^-}S^*_{R}+\Phi^*_{m,w^n}\Phi^*_{n,R}-\Phi^*_{m,R}\frak{g}^{-R}\right)=0,\\
\frak{m}_{mn}&=&(D_s\Upsilon)_{mn,R}-\left(\Upsilon_{mm.w^+}S_{R}+\Upsilon_{mn,w^-}+\Upsilon_{mn,w^k}\Phi_{k,R}\right)\nonumber\\
&&-\left(\Phi_{m,w^+}\Phi_{n,R}+\Phi_{m,w^-}\Phi^*_{n,R}+\Phi_{m,w^k}\Upsilon_{nk,R}\right)\nonumber\\
&&-\left(\Phi_{n,w^+}\Phi_{m,R}+\Phi_{n,w^-}\Phi^*_{m,R}+\Phi_{n,w^k}\Upsilon_{mk,R}\right)\nonumber\\
&&+\Upsilon_{mn,R}\frak{g}^{+R}+\Phi_{m,R}\frak{g}^{nR}+\Phi_{n,R}\frak{g}^{mR}=0,\\
\frak{m^*}_{mn}&=&(D_{s^*}\Upsilon)_{mn,R}-\left(\Upsilon_{mm.w^-}S^*_{R}+\Upsilon_{mn,w^+}+\Upsilon_{mn,w^k}\Phi^*_{k,R}\right)\nonumber\\
&&-\left(\Phi^*_{m,w^-}\Phi^*_{n,R}+\Phi^*_{m,w^+}\Phi_{n,R}+\Phi^*_{m,w^k}\Upsilon_{nk,R}\right)\nonumber\\
&&-\left(\Phi^*_{n,w^-}\Phi^*_{m,R}+\Phi^*_{n,w^+}\Phi_{m,R}+\Phi^*_{n,w^k}\Upsilon_{mk,R}\right)\nonumber\\
&&+\Upsilon_{mn,R}\frak{g}^{-R}+\Phi^*_{m,R}\frak{g}^{nR}+\Phi^*_{n,R}\frak{g}^{mR}=0.\label{w52ap}
\end{eqnarray}

\end{document}